\documentclass[12pt]{iopart}
\def\be{\begin{equation}}
\def\ee{\end{equation}}
\def\bea{\begin{eqnarray}}
\def\eea{\end{eqnarray}}
\def\nn{\nonumber}

\def\Ga{\Gamma}

\def\la{\lambda}

\def\si{\sigma}

\begin{document}

\title{The energy of the universe in teleparallel gravity}

\author{T Vargas}
\address{Instituto de F\'{\i}sica Te\'orica\\
Universidade Estadual Paulista \\
Rua Pamplona 145 \\
01405-900\, S\~ao Paulo SP \\
Brazil}

\begin{abstract}
The teleparallel versions of the Einstein and the Landau-Lifshitz
energy-momentum complexes of the gravitational field are obtained. 
By using these complexes, the total energy of the universe, which includes 
the energy of both the matter and the gravitational fields, is then 
obtained. It is shown that the total energy vanishes independently 
of both the curvature parameter and the three dimensionless coupling 
constants of teleparallel gravity.

\end{abstract}


\maketitle

\section {Introduction}

The notion of energy and/or momentum localization of the gravitational field 
is one of the oldest and most controversial problems of the  general theory of
relativity~\cite{mtw}. Following the energy-momentum pseudotensor of
Einstein~\cite{tra}, several other prescriptions have been introduced, leading
to a great variety of expressions for the energy-momentum pseudotensor of the
gravitational field~\cite{lal}. These pseudotensors are not covariant objects
because they inherently depend on the reference frame, and thus cannot
provide a truly physical local gravitational energy-momentum density. The
physical origin of this difficulty lies in the principle of equivalence,
according to which a gravitational field should not be detectable at a point. 
Consequently, the pseudotensor approach has been largely questioned, although
never abandoned. A quasilocal approach has been proposed and has been widely
accepted~\cite{bryo}. However, it has been shown by Bergqvist~\cite{ber}, that
there are an  {\it infinite number} of quasilocal expressions for the
gravitational field. Still in this context, it has been shown recently by Chang
and Nester~\cite{nest} that every energy energy-momentum  pseudotensor
can be associated with a particular Hamiltonian boundary term. In this sense
it is usually said that the quasilocal energy-momentum rehabilitates the
pseudotensor approach. A natural question then arises: What role does this
gravitational energy density play in the description of the total energy of the
universe? For example, during inflation the vacuum energy driving the
accelerated expansion of the universe, and which was responsible for the
creation of radiation and matter in the universe, is drawn from the
energy of the gravitational field~\cite{prig}. Such transition of gravitational 
energy into of a ``cosmological fluid'' must face problems of localization on 
account of the problems discussed above. Despite this difficulty, there has 
been several attempts~\cite{coop,ros,other} to calculated the total energy
of the expanding universe. In one of these attempts the Einstein
energy-momentum pseudotensor has been used to represent the gravitational
energy~\cite{ros}, which led to the result that the total energy of a closed
Friedman-Robertson-Walker (FRW) universe is zero. In another
attempt, the symmetric pseudotensor of Landau-Lifshitz has 
been used~\cite{other}. In~\cite{baxu}, the total energy of the 
anisotropic Bianchi models has been calculated using different pseudotensors, leading
to a similar result. Several other attempts, using Killing
vectors, or using the conservation law in the vierbein formulation, also led to the
same conclusion. Recently~\cite{facoo}, it has been shown that open, or
critically open FRW universes, as well as Bianchi models evolving into de Sitter
spacetimes also have zero total energy.       

An alternative approach to gravitation is the so called teleparallel 
gravity~\cite{M}, which corresponds to a gauge theory for the translation
group based on the Weitzenb\"ock geometry~\cite{we}. In this theory, 
gravitation is attributed to torsion~\cite{xa}, which plays the role of a 
force~\cite{pe}, and the curvature tensor vanishes identically. The
fundamental field is represented by a nontrivial tetrad field, which gives 
rise to the metric as a by-product. The translational gauge potentials appear
as the nontrivial part of the tetrad field, and thus induces on spacetime a
teleparallel structure which is directly related to the presence of the
gravitational field. The interesting point of teleparallel gravity is that,
due to its gauge structure, it can reveal a more appropriate approach to
consider some specific problem. This is the case, for example, of the
energy-momentum problem, which becomes more transparent when considered from
the teleparallel point of view. In fact, it has been shown recently that the
energy-momentum gauge current associated to the teleparallel gravity is a
true tensor that reduces to the M{\o}ller's canonical energy-momentum density
of the gravitational field when returning to the geometrical
approach~\cite{andrad}.

By working then in the context of teleparallel gravity, the basic purpose of
this paper will be to obtain the teleparallel version of both the Einstein and
Landau-Lifshitz energy-momentum complexes. As an application, the energy of a
FRW universe will be calculated, which includes the energy of matter,
as well as the energy of the gravitational field. It will be shown that the
total energy vanishes independently of the curvature parameter and the three 
dimensionless coupling constants of teleparallel gravity. We will proceed 
according to the following scheme. In section~2, we review the main features of 
teleparallel gravity, and obtain the teleparallel versions of Einstein and 
Landau-Lifshitz complexes. In section~3, it is obtained the tetrad field, the 
non-zero components of the Weitzenb\"ock connection, the torsion tensor, and the 
superpotentials for the FRW universe in Cartesian coordinates. Discussions and 
conclusions are presented in section~4.
  
\section {Teleparallel gravity}

In teleparallel gravity, spacetime is represented by the Weitzenb\"{o}ck
manifold $W^{4}$ of distant parallelism. This gravitational theory naturally
arises within the gauge approach based on the group of the spacetime
translations. Accordingly, at each point of this manifold, a gauge
transformation is defined as a local translation of the tangent-space
coordinates,
\[
x^{a}\rightarrow x'^a  = x^{a}+b^{a}
\]
where $b^{a} = b^{a}(x^{\mu})$ are the transformation parameters.
For an infinitesimal transform\-ation, we have
\[
\delta x^a = \delta b^{c} P_{c} \, x^a,
\]
with $\delta b^{a}$ the infinitesimal parameters, and $P_{a}=\partial_{a}$
the generators of translations. Denoting the translational gauge potential by
$A^{a}{}_{\mu}$, the gauge covariant derivative for a scalar field
$\Phi(x^{\mu})$ reads~\cite{pe} 
\be
D_{\mu}\Phi=h^{a}{}_{\mu}\partial_{a}\Phi,
\label{drc}
\ee
where
\be
h^{a}{}_{\mu} = \partial_{\mu}x^{a} + A^{a}{}_{\mu}
\label{tetra}
\ee
is the tetrad field, which satisfies the orthogonality condition
\be
h^{a}{}_{\mu}\,h_{a}{}^{\nu}=\delta^{\nu}_{\mu} .
\label{ort}
\ee
This nontrivial tetrad field induces a teleparallel structure on spacetime
which is directly related to the presence of the gravitational field, and the 
Riemannian metric arises as
\be
g_{\mu \nu} = \eta_{a b} \; h^a{}_\mu \; h^b{}_\nu \; .
\label{met}
\ee
In this theory, the fundamental field is a nontrivial tetrad, which gives
rise to the metric as a by-product. The parallel transport of the tetrad
$h^{a}{}_{\mu}$ between two neighbouring points is encoded in the covariant
derivative
\be
\nabla_{\nu}h^{a}{}_{\mu}=\partial_{\nu}h^{a}{}_{\mu}
-\Gamma ^{\alpha}{}_{\mu \nu}h^{a}{}_{\alpha},
\label{nt}
\ee
where $\Gamma^{\alpha}{}_{\mu \nu}$ is the  Weitzenb\"ock connection.
Imposing the condition that the tetrad be parallel transported in the
Weitzenb\"ock space-time, we obtain
\[
\nabla_{\nu}h^{a}{}_{\mu}=\partial_{\nu}h^{a}{}_{\mu} 
- \Gamma^{\alpha}{}_{\mu \nu}h^{a}{}_{\alpha}\equiv 0.
\label{trt}
\]
This is the condition of absolute parallelism, or teleparallelism~\cite{xa}.
It is equivalent to
\be
\Gamma ^{\alpha}{}_{\mu \nu}=h_{a}{}^{\alpha}\partial_{\nu}h^{a}{}_{\mu}
\ee
which gives the explicit form of the Weitzen\-b\"ock connection in
terms of the tetrad, and
\be
T^{\rho}{}_{\mu \nu}=\Gamma ^{\rho}{}_{\nu \mu}-\Gamma ^{\rho}{}_{\mu \nu}
\ee
is the torsion of the Weitzenb\"ock connection. As we already remarked, 
the curvature of the Weitzenb\"ock connection vanishes identically as a
consequence of absolute parallelism.   

The action of teleparallel gravity in the presence of matter is given by
\be
S = \frac{1}{16 \pi G} \int d^{4}x \, h\,S^{\la \tau \nu} \; T_{\la \tau \nu}+
\int d^{4}x \, h\,{\mathcal L_{M}} \label{lag}
\ee
where $h = \det(h^a{}_\mu$), $\mathcal L_{M}$ is the Lagrangian of the matter
field, and $S^{\la \tau \nu}$ is the tensor
\be
S^{\la \tau \nu} = c_1T^{\la \tau \nu}+\frac{c_2}{2}\left(T^{\tau \la \nu} -
T^{\nu \la \tau}\right) +\frac{c_3}{2}\left(g^{\la \nu} \;
T^{\si \tau}{}_{\si} - g^{\tau \la} \;T^{\si \nu}{}_{\si} \right).\label{2.32a}
\ee
with $c_1$, $c_2$, and $c_3$ the three dimensionless coupling constants of
teleparallel gravity~\cite{xa}. For the specific choice
\be
c_1=\frac{1}{4}, \quad c_2=\frac{1}{2}, \quad c_3=-1,
\label{par}
\ee
teleparallel gravity yields the so called teleparallel equivalent of general
relativity.

By performing variation in (\ref{lag}) with respect to $h^{a}{}_{\mu}$,
we get the teleparallel field equations,
\be
\partial_\sigma(h S_{\lambda}{}^{\tau \sigma}) - 4 \pi G\; 
(ht^{\tau}{}_{\la}) = 4 \pi G\;h\; T^{\tau}{}_{\la},     \label{eqc}
\ee
where
\be
t^{\tau}{}_{\la} = \frac{1}{4 \pi G}h\Gamma^{\nu}{}_{\sigma \lambda}
S_{\nu}{}^{\tau \sigma} - \delta^{\tau}{}_{\la}\mathcal L_{G}
\ee
is the energy-momentum pseudotensor of the gravitational field~\cite{andrad}. 
Rewriting the teleparallel field equations in the form
\be
h(t^{\tau}{}_{\la} + T^{\tau}{}_{\la}) = \frac{1}{4 \pi G} 
\partial_\sigma(h S_{\lambda}{}^{\tau \sigma}),     \label{pse}
\ee
as a consequence of the antisymmetry of $S_{\lambda}{}^{\tau \sigma}$ in 
the last two indices, we obtain immediately the conservation law 
\be
\partial_{\tau}[h (t^{\tau}{}_{\la}+T^{\tau}{}_{\la})] = 0.
\ee
For the particular choice (\ref{par}) of the parameters, on account of the
identity
\be
\partial_\sigma(h S_{\lambda}{}^{\tau \sigma}) - 4 \pi G\; 
(ht^{\tau}{}_{\la}) \equiv  \frac{h}{2}\;\left(R^{\tau}_{\la} -
\frac{1}{2}\delta^{\tau}{}_{\la} R \right),
\ee
the teleparallel field equation is the same as Einstein's equation~\cite{pe}.
Using this equivalence, as well as Eq.\ (\ref{pse}), we find that 
$h S_{\lambda}{}^{\tau \sigma}=U_{\lambda}{}^{\tau \sigma}$
coincides with Freud's superpotential. Consequently, $t^{\tau}{}_{\la}$ is
nothing but the teleparallel version of Einstein's gravitational
energy-momentum pseudotensor. This superpotential and the Lagrangian $\mathcal
L_{G}$ of the gravitational field are related by
\be
U_{\lambda}{}^{\tau \sigma} = 4\pi Gh^{a}{}_{\la}\frac{\partial\mathcal L_{G}}
{\partial(\partial_{\sigma}h^{a}{}_{\tau})}\, .  
\ee
Equation (\ref{pse}), therefore, can be rewritten as 
\be
h \; \mathcal T_{E}{}^{\tau}{}_{\la} = \frac{1}{4 \pi G} 
\partial_\sigma(U_{\lambda}{}^{\tau \sigma}),  \label{ein}
\ee
where $\mathcal T_{E}{}^{\tau}{}_{\la}$ is the Einstein energy-momentum
complex, which is given by the divergence of the Freud's superpotential. The
Bergmann-Thompson energy-momentum complex, on the other hand, is 
\be
h\;\mathcal T_{BT}{}^{\mu \tau} = \frac{1}{4 \pi G} 
\partial_\sigma(g^{\mu \lambda} U_{\lambda}{}^{\tau \sigma}), \label{bt}
\ee
whereas the Landau-Lifshitz complex is
\be
h\;\mathcal T_{LL}{}^{\mu \tau} = \frac{1}{4 \pi G} 
\partial_\sigma(h\;g^{\mu \lambda} U_{\lambda}{}^{\tau \sigma}). \label{ll}
\ee
For anyone of the cases, we have the relation,
\be
P_{\la} = \int_{\Omega}h \mathcal T{}^{0}{}_{\la} \; d^{3}x   \label{ei}
\ee
where $P_{0}$ is the energy, while $P_{i}$ stand for the four-momentum 
components and the integration hypersurface $\Omega$ is defined by 
$x^0 = t =$ constant. We remark that, for our purposes, it is not necessary to 
know the explicit form of the Einstein and Landau-Lifshitz gravitational 
energy-momentum pseudotensors. Instead, it is sufficient the relation between 
these complexes and their corresponding superpotentials, given by Eqs.\
(\ref{ein}) and (\ref{ll}).  

\section {The teleparallel homogeneous isotropic type solution}
 
The line element of the homogeneous isotropic FRW universe is given by
\be
ds^{2} = dt^{2} - \frac{a(t)^2}{(1+\frac{kr^2}{4})}(dr^{2} + r^{2}d\theta^{2}+
r^{2}sin^{2}{\theta}d\phi^{2}),\label{dsi}
\ee
where $a(t)$ is the time-dependent cosmological scale factor, and $k$ is
the curvature parameter $k = 0, \pm 1$. As remarked in Ref.~\cite{rovi}, it is
important  to work with Cartesian coordinates, as other coordinates may lead to
non-physical values for the pseudotensor $t^{\tau}{}_{\la}$. Transforming,
therefore, from polar to Cartesian coordinates, the FRW line element
(\ref{dsi}) becomes
\be
ds^{2}= dt^{2}- \frac{a(t)^2}{(1+\frac{kr^2}{4})}(dx^{2}+dy^{2}+dz^{2}).
\label{dsi1}
\ee
Using the relation (\ref{met}), we obtain the tetrad components:
\be
\label{te1}
h^{a}{}_{\mu}= {\rm diag} \left(1,\;-\frac{a(t)}{1+\frac{kr^2}{4}},
\;-\frac{a(t)}{1+\frac{kr^2}{4}},\;-\frac{a(t)}{1+\frac{kr^2}{4}}\right).
\ee
Its inverse is
\be
\label{te2}
h_{a}{}^{\mu}= {\rm diag} \left(1,\;-\frac{1+\frac{kr^2}{4}}{a(t)},
\;-\frac{1+\frac{kr^2}{4}}{a(t)},\;-\frac{1+\frac{kr^2}{4}}{a(t)}\right).
\ee
One can easily verify that the relations (\ref{ort}) and (\ref{met}) 
between $h^{a}{}_{\mu}$ and $h_{a}{}^{\mu}$ are satisfied.

From Eqs.(\ref{te1}) and (\ref{te2}), we can now construct the Weitzenb\"ock
connection, whose nonvanishing components are found
\bea
\Ga^{x}{}_{xt} = \Ga^{y}{}_{yt} = \Ga^{z}{}_{zt}= \frac{\dot{a}(t)}{a(t)}, \nn \\ 
\Ga^{x}{}_{x x} =\Ga^{y}{}_{y x}=\Ga^{z}{}_{z x}=-\frac{kx}{2(1+\frac{k r^2}{4})}, \nn \\
\Ga^{x}{}_{x y} =\Ga^{y}{}_{y y}=\Ga^{z}{}_{z y}=-\frac{ky}{2(1+\frac{k r^2}{4})}, \nn \\
\Ga^{x}{}_{x z} =\Ga^{y}{}_{y z}=\Ga^{z}{}_{z z}=-\frac{kz}{2(1+\frac{k r^2}{4})}. \nn
\eea
where a dot denotes a derivative with respect to the time $t$. The 
corresponding non-vanishing torsion components are:
\bea
T^{x}{}_{tx} = T^{y}{}_{ty} = T^{z}{}_{tz} = \frac{\dot{a}(t)}{a(t)}, \nn \\
T^{x}{}_{x y} =T^{z}{}_{z y}=\frac{ky}{2(1+\frac{k r^2}{4})}, \nn \\
T^{x}{}_{x z} =T^{y}{}_{y z}=\frac{kz}{2(1+\frac{k r^2}{4})}, \nn \\
T^{y}{}_{y x} =T^{z}{}_{z x}=\frac{kx}{2(1+\frac{k r^2}{4})} \nn
\eea
Now, the non-zero components of the tensor $S_{\nu}{}^{\sigma \tau}$ read
\bea
S_{t}{}^{tx} = S_{y}{}^{yx} = S_{z}{}^{zx}=-\frac{kx}{2 a^2} \left(1+\frac{k r^2}{4}\right)
\left(c_1+\frac{c_2}{2}+c_3\right),\nn \\
S_{t}{}^{ty} = S_{x}{}^{xy} = S_{z}{}^{zy} =-\frac{ky}{2 a^2} \left(1+\frac{k
r^2}{4} \right) \left( c_1+\frac{c_2}{2}+c_3 \right),\nn \\
S_{t}{}^{tz} = S_{x}{}^{xz} = S_{y}{}^{yz} =-\frac{kz}{2 a^2} \left( 1+\frac{k r^2}{4}
\right) \left(c_1+\frac{c_2}{2}+c_3 \right), \nn \\
S_{x}{}^{xt} = S_{y}{}^{yt} = S_{z}{}^{zt}= -\frac{\dot{a}(t)}{a(t)}
\left( c_1+\frac{c_2}{2}+c_3 \right). \nn
\eea
In more compact form, they are
\be
U_{t}{}^{ \sigma \tau}\equiv hS_{t}{}^{\sigma \tau}=-\frac{1}{2}\left(\delta^{\sigma}
_{t}\delta^{\tau}_{i}-\delta^{\tau}_{t} \delta^{\sigma}_{i}\right) \; \frac{ka(t)x^{i}}
{(1+\frac{k r^2}{4})} \; \left(c_1+\frac{c_2}{2}+c_3 \right).
\label{po}
\ee

Let us now calculate the total energy of the FRW universe at the instant 
$x^0=t=$ constant, which is given by the integral over the space section. As 
in Ref~\cite{ros}, we carry out the integration in polar coordinates. For the Einstein
energy-momentum complex, using Eqs.~(\ref{ein}), (\ref{ei}) and (\ref{po}), we have 
\bea
E &=&k \frac{a(t)}{G} \; (c_1+\frac{c_2}{2}+c_3)\left[\frac{1}{2}
\int\limits^\infty_0\frac {kr^4 dr}{(1+\frac{kr^2}{4})^3}-\frac{3}{2}
\int\limits^\infty_0\frac{r^2 dr} {(1+\frac{kr^2}{4})^2}\right].
\label{ek}
\eea
For a closed universe $(k=+1)$, the energy is   
\bea
E &=& \frac{a(t)}{G} \; (c_1+\frac{c_2}{2}+c_3)\left[\frac{1}{2} \int\limits^\infty_0\frac
{r^4 dr}{(1+\frac{r^2}{4})^3}-\frac{3}{2} \int\limits^\infty_0\frac{r^2 dr}
{(1+\frac{r^2}{4})^2}\right] \nn \\
&=& \frac{a(t)}{G} \; (c_1+\frac{c_2}{2}+c_3)
\left[\frac{1}{2}6\pi-\frac{3}{2}2\pi \right]=0,
\eea
which is the same as Rosen's results~\cite{ros}. For an open universe $(k=-1)$, we
obtain 
\bea
E &=& \frac{a(t)}{G} \; (c_1+\frac{c_2}{2}+c_3)\left[\frac{3}{2}
\int\limits^\infty_0\frac {r^2 dr}{(1-\frac{r^2}{4})^2}+\frac{1}{2}
\int\limits^\infty_0\frac{r^4 dr} {(1-\frac{r^2}{4})^3}\right] \nn \\
 &=& \frac{a(t)}{G} \; (c_1+\frac{c_2}{2}+c_3)
\left[\frac{3}{2}(\mp 2\pi i)+\frac{1}{2}(\pm 6\pi i)\right]=0. 
\eea
Finally, for the spatially flat universe $(k=0)$, Eq.~(\ref{ek}) gives again $E=0$. 

Now, for the Landau-Lifshitz complex, using (\ref{ll}), (\ref{ei}) and (\ref{po}), the
total energy is found to be
\bea
E &=&k \frac{a^4(t)}{G} \; (c_1+\frac{c_2}{2}+c_3)\left[\frac{5}{4}
\int\limits^\infty_0\frac {kr^4 dr}{(1+\frac{kr^2}{4})^6}-\frac{3}{2}
\int\limits^\infty_0\frac{r^2 dr} {(1+\frac{kr^2}{4})^5}\right]. 
\label{llk}
\eea
The energy of a closed universe $(k=+1)$ is 
\bea
E &=& \frac{a^4(t)}{G} \; (c_1+\frac{c_2}{2}+c_3)\left[\frac{5}{4}
\int\limits^\infty_0\frac {r^4 dr}{(1+\frac{r^2}{4})^6}-\frac{3}{2}
\int\limits^\infty_0\frac{r^2 dr} {(1+\frac{r^2}{4})^5}\right] \nn \\
&=&  \frac{a^4(t)}{G} \; (c_1+\frac{c_2}{2}+c_3)
\left[\frac{5}{4}\frac{3\pi}{16}-\frac{3}{2}\frac{5\pi}{32}\right]=0, 
\eea 
which confirms the Johri {\it et al.} results~\cite{other}. For an open universe
$(k=-1)$,
\bea
E &=& \frac{a^4(t)}{G} \; (c_1+\frac{c_2}{2}+c_3)\left[\frac{3}{2}
\int\limits^\infty_0\frac {r^2
dr}{(1-\frac{r^2}{4})^5}+\frac{5}{4}\int\limits^\infty_0\frac{r^4 dr}
{(1-\frac{r^2}{4})^6}\right] \nn \\  &=& \frac{a^4(t)}{G} \;  (c_1+\frac{c_2}{2}+c_3)
\left[\frac{3}{2} \left(\mp \frac{5\pi i}{32} \right) + \frac{5}{4} \left(\pm \frac{3\pi
i}{16} \right)\right]=0.   
\eea
And finally, for the spatially flat universe $(k=0)$, Eq.~(\ref{llk}) gives again
$E=0$.

\section {Final Remarks}

Working in the context of teleparallel gravity, we have calculated in the
total energy of the FRW universe, which includes the energy of the matter and that of
the gravitational field. In order to compute the gravitational part of the energy, we
have considered the teleparallel version of both Einstein and Landau-Lifshitz
energy-momentum complexes. Our basic result is that the total energy vanishes whatever
be the pseudotensor used to describe the gravitational energy. It is also independent
of both the curvature parameter and the three teleparallel dimensionless coupling
constants. It is valid, therefore, not only in the telaparallel equivalent of general
relativity, but also in any teleparallel model.

Finally, it is important to remark that, for an open FRW universe, the result that the
total energy vanishes is quite {\it unexpected}. In fact, as the open FRW universe is
an infinite  spacetime filled with matter and gravitational field, our common sense
would point to an infinite total energy. However, the result obtained is that the
total energy for this universe is zero. We can thus conclude that the gravitational
energy exactly cancels out the matter energy, and that this cancellation is independent
of the curvature parameter.

\ack
The author would like to thank J. G. Pereira for fruitful discussions, for reading 
and improving the English in the manuscript. He would like also to thank FAPESP for 
financial support.

\end{document}